\documentclass[12pt,pra,aps,amssymb,amsfonts,amsmath,tightenlines]{revtex4}
\usepackage{graphicx}
\usepackage{dsfont}
\usepackage{pxfonts}
\usepackage{overpic}

\newcommand{\half}{\mbox{$\textstyle \frac{1}{2}$}}
\newcommand{\quarter}{\mbox{$\textstyle \frac{1}{4}$}}
\newcommand{\octa}{\mbox{$\textstyle \frac{1}{8}$}}

\newtheorem{definition}{Definition} 
 
\newtheorem{prop}{Proposition}

\begin{document}

\title[Quantum State Reduction]
{Quantum State Reduction}

\author{Dorje C. Brody$^{1,2}$ and Lane P. Hughston$^{1}$}

\affiliation{$^1$Department of Mathematics, Brunel University London, Uxbridge UB8 3PH, UK  \\ 
              $^2$Department of Optical Physics and Modern Natural Science, \\ 
St Petersburg National Research University of Information Technologies, Mechanics and Optics, 
49 Kronverksky Avenue, St Petersburg 197101, Russia}

\date{\today}

\begin{abstract}
\noindent 
We propose an energy-driven  stochastic master equation for the density matrix as a dynamical model for quantum state reduction. In contrast, most previous studies of state reduction have considered stochastic extensions of the Schr\"odinger equation, and have introduced the density matrix as the expectation of the random pure projection operator associated with the evolving state vector. After working out properties of the reduction process we construct a general solution to the energy-driven stochastic master equation. The solution  is obtained by the use of nonlinear filtering theory and takes the form of a completely positive stochastic map.

\vspace{-0.1cm}
\begin{center}
{\scriptsize {\bf Keywords: quantum mechanics, collapse of wave function, measurement problem, density matrix,  
\\ \vspace{-0.2cm} 
master equation, stochastic analysis, nonlinear filtering
} }
\end{center}

\end{abstract}


\maketitle

\section{Introduction}
\label{introduction}

 
\noindent Many physicists have expressed the view that quantum mechanics needs to be modified 
 to provide a mechanism for ``collapse of the wave function" 
(Pearle 1976, Penrose 1986, Bell 1987, Diosi 1989, Ghirardi 2000, Adler 2003a,  Weinberg 2012). Among the ways forward that have been proposed, perhaps the most fully developed, at least from a mathematical point of view, are the so-called stochastic models for state reduction, in connection with which there is now a substantial body of literature. In such models, the quantum system is usually taken to be in a pure state, represented by a vector in Hilbert space, evolving as a stochastic process. The state of the system evolves randomly in such a way that it eventually approaches an eigenstate of a preferred observable, such as position or energy. In the situation where the reduction is to a state of definite energy,
which is the case 
that will concern us here, the setup is as follows. 
The Hilbert space $\mathcal H$ is taken to be of finite dimension 
$N$, and the state-vector process $\{|\psi_t\rangle\}_{t \geq 0}$ is assumed to satisfy an Ito-type stochastic differential equation of the form
\begin{eqnarray}
\label{pure dynamics}
{\rm d}|\psi_t\rangle &=& -{\rm i} \, \hbar^{-1} {\hat H} |\psi_t\rangle \, {\rm d}t
- \octa \sigma^2({\hat H}-H_t)^2 |\psi_t\rangle \, {\rm d}t + \half
\sigma ({\hat H}-H_t)|\psi_t\rangle \, {\rm d}W_t . 
\end{eqnarray}
Here $\{W_t\}_{t \geq0}$ is a standard Brownian motion, and
$|\psi_t\rangle \in \mathcal H$ is the state vector at time $t$. The initial state vector
$|\psi_0\rangle$ is an input of the model.  We write
\begin{eqnarray}
\label{energy process}
H_t = \frac{\langle{\psi}_t| \,{\hat H} \, |\psi_t\rangle}
{\langle{\psi}_t|\psi_t\rangle}  
\end{eqnarray}
for the expectation value of the Hamiltonian operator ${\hat H}$ in the
state $|\psi_t\rangle$.  The reduction parameter $\sigma$, which has dimensions such that 
\begin{eqnarray}
\sigma^2 \approx [\rm {energy}]^{-2} [\rm {time}]^{-1}, 
\end{eqnarray}   
determines the characteristic timescale $\tau_R$ associated with the reduction of the state, which is  of the order $ \tau_R  \approx 1/(\sigma \, \Delta H)^2$,
where $\Delta H$ is the initial uncertainty of the energy. Thus a state with high initial energy uncertainty has a shorter characteristic reduction timescale than a state with low energy uncertainty. After a few multiples of  $\tau_R$, the system will be nearly in an eigenstate of energy. The determination of $\sigma$ is an empirical matter.  
One intriguing possibility suggested by a number of authors (Karolyhazy 1966,  Karolyhazy et al 1986,  Penrose 1986, 1996, Diosi 1989,  Percival 1994, Hughston 1996) is that state reduction is determined in some way by gravitational phenomena. In that case we might suppose that $\sigma$ is given by a relation of the form $\sigma^2 \approx {E_P}^{-2} {T_P}^{-1}$, where $E_P$ is the Planck energy and $T_P$ is the Planck time, and hence of the order
\begin{eqnarray}
\sigma^2  \approx \sqrt{G\hbar^{-3}c^{-5}}.
\end{eqnarray}
A surprising feature of this expression is that the large numbers associated with the various physical constants cancel out, and we are left with a reduction timescale that is in principle observable in the laboratory, given by
\begin{eqnarray}
\tau_{R} \approx \left( \frac{2.8\,{\rm MeV}} {\Delta H}\right)^2 {\rm s}. 
\end{eqnarray}
Going forward, we shall not make any specific assumptions regarding the magnitude of the reduction parameter. Nevertheless, to get a feeling for the numbers involved, we note that the binding energies per nucleon of low mass nuclei are of the order of 1.1 MeV for the deuteron, 2.6 MeV for ${\rm He^3}$, and  7.1 MeV for ${\rm He^4}$. Since the fusion reactions leading to the production of such nuclei are essential in normal stellar evolution, it is not unreasonable to suppose that some form of observer-free ``objective" state reduction is involved in the process, and that gravitational effects play a role as well. 

No attempt will be made to review the extensive literature of dynamical collapse models, 
 of which the energy-driven model described above is an example, or to discuss in any detail the relative merits of the various models that have been proposed. See Bassi \& Ghirardi (2003), Bassi (2007), Pearle (2007, 2009), Bassi et al (2013), Ghirardi (2016) for surveys. For aspects of the energy-driven models, we refer the reader to
Gisin (1989), Ghirardi et al (1990), Percival (1994, 1998), Hughston (1996), Pearle (1999, 2004), Adler \& Horwitz (2000),  Adler \& Mitra (2000), Adler et al (2001), Brody \& Hughston (2002a,b, 2005, 2006),  Adler (2003a,b, 2004), Brody et al (2003, 2006),  Gao (2013), Meng\"ut\"urk (2016).  Adler (2002), in an empirical study of energy-driven models, concludes thus: 
\begin{quote}
Our analysis supports the suggestion that a measurement takes place when the different outcomes are characterized by sufficiently distinct environmental interactions for the reduction process to be rapidly driven to completion.
\end{quote}
Although other collapse models have been considered at length in the literature, including, for example, the GRW model (Ghirardi, Rimini \& Weber 1986) and so-called {\sl continuous spontaneous localization} (CSL) models (Diosi 1989, Pearle 1989, Ghirardi, Pearle \& Rimini 1990), the energy-driven reduction models stand out, in our view, on account of (a) their parsimonious mathematical structure, and (b) the fact that they are universal. 
By ``universal", we mean applicable to any quantum system.   We point out that energy-driven models maintain the conservation of energy in a well-defined probabilistic sense, as an extension of the Ehrenfest theorem, whereas models driven by observables that do not commute with the energy, such as position,  do not conserve energy (Pearle 2000, Bassi, Ippoliti \& Vacchini 2005).   Furthermore, energy-driven models give the Born rule and the L\"uders projection postulate as exact results (Adler \& Horwitz 2000, Adler et al 2001, Adler 2003b), whereas other models do not.  For these reasons we emphasize here the role of energy-driven models. This is not to say that  energy-driven models are the only ones to be taken seriously. But if one wishes to propose a stochastic reduction model that it is applicable to {\sl any} nonrelativistic system, without qualification, {\sl including finite dimensional systems}, then  it must be an energy-driven model. 

In that case, is the dynamics necessarily of the form (\ref{pure dynamics}) given above? Clearly not, since, for a start, one could consider the possibility that other forms of noise than Brownian motion act as a basis for the stochastic dynamics of the state, and indeed there is a sizable literature dealing with dynamical reduction models based on other types of noise. To keep the discussion focussed, we stick here with models based on Brownian motion, though in the final section of the paper we comment briefly on a generalization to models based on L\'evy noise. One might also introduce time-dependent coupling (Brody \& Hughston 2005, 2006, Brody et al 2006, Meng\"ut\"urk 2016), which offers an approach to the ``tail problem" (Shimony 1990, Pearle 2009). Again, we pass over such considerations for the present. 

There is, however, an important aspect of the dynamical equation (\ref{pure dynamics}) that seems to build in what might be viewed as an unnecessary assumption, even if one accepts the principle that reduction must be energy driven, and even if one narrows the scope to models based on a Brownian filtration. This concerns the  issue of what constitutes a ``state" in quantum mechanics. The physics community seems to be divided on the matter. 
It is worth recalling that in von Neumann's highly influential 1932 book, the term ``state" is reserved for {\sl pure} states, and the statistical operator is introduced to describe {\sl mixtures}. He introduces the notion of a statistical ensemble, corresponding to a countable collection of quantum systems, each of which is in a pure state, and he distinguishes two cases. In the first case, the individual systems of the ensemble can be in different states, and the statistical operator is determined by their relative frequencies. In the second case, which he calls a homogeneous ensemble, the various individual systems are in the same state. The statistical operator for a homogeneous ensemble is identical to the state of any one of its elements, and takes the form of a pure projection operator. 

In his consideration of statistical ensembles von Neumann (1932) was motivated in part by the frequentist theories of von Mises (1919, 1928). In particular, von Neumann identifies his concept of ensemble with von Mises's idea of a ``Kollektiv"  (random sequence): 
\begin{quote}
Such ensembles, called collectives, are in general necessary for establishing probability theory as the theory of frequencies. They were introduced by Richard von Mises, who discovered their meaning for probability theory, and who built up a complete theory on this foundation. 
\end{quote}
According to von Mises, ``Erst das Kollektiv, dann die Wahrscheinlichkeit".  At about the same time that these developments were under way, Kolmogorov (1933) revolutionized classical probability theory by giving it a set-theoretic foundation and providing it with a subtle measure-theoretic definition of conditional expectation that allows one to handle in a satisfactory way the  logical issues associated with conditioning on events of probability zero. 
The mathematics community took on board Kolmogorov's innovations, and success followed success, with the introduction of many further new ideas, including, among others, martingales, stochastic calculus, and nonlinear filtering. 
Von Mises's theory, despite its attractive features, was eventually dropped by mathematicians, even though the ensemble concept (and elements of the frequentist thinking underpinning it) has been kept alive by physicists, and is still  taught to students (Isham 1995, for instance, gives a good treatment of the relevant material). See van Lambalgen (1999) for a rather detailed discussion of where von Mises's ideas stand today. 
It appears that the more general use of the term ``state"  (to include mixed as well as pure states) was introduced by Segal (1947), in his postulates for general quantum mechanics. Segal's point of view was adopted by Haag \& Kastler (1964), and also by Davies (1976), who says: 
\begin{quote}
The states are defined as the non-negative trace class operators of trace one, elsewhere called mixed states or density matrices. 
\end{quote}
If the matter were purely one of terminology, there would be no point in worrying about it very much. The problem is that in the language physicists use there can be assumptions that are implicit in the choice of words, and these in turn can guide the direction of the subject as it moves forward. The issue of what exactly constitutes a ``state" is such a case. 

The point that concerns us here is that most of the models that have been developed in detail in the collapse literature treat the quantum system as a randomly evolving {\it pure} state. This point of view is represented, for example, in Ghirardi, Pearle \& Rimini (1990) in the context of their development of the CSL model, where we find the following succinct account of their stance on the matter: 
\begin{quote}
The theory discussed here allows one to describe naturally quantum measurement processes by dynamical equations valid for all physical systems. It is worthwhile repeating, that, in this theoretical scheme, any member of the statistical ensemble has at all times a definite wave function. As a consequence, the wave function itself can be interpreted as a real property of a single closed physical system.
\end{quote}
The emphasis placed on the role of pure states reflects a view held by many physicists that pure states should be treated as being fundamental. See, for example, Penrose (2016), who argues persuasively concerning the preferred status of pure states. According to this view, which, as we have indicated, is generally in line with that of von Neumann (1932), individual systems are represented by pure states.  Physicists are likewise divided on the issue of the status of statistical ensembles. Are they essential to the theory? Mielnik (1974) offers the following:  
\begin{quote}
It is an old question whether the formalism of quantum theory is adequate to describe the properties of single systems. What is verified directly in the most general quantum experiment are rather the properties of statistical ensembles. 
\end{quote}
Although our brief remarks cannot do justice to the deep insights of the authors mentioned above, one will be impressed by the diversity of opinion held by physicists on the nature of quantum states and the role of statistical ensembles.  It should be emphasized, nevertheless, that, as far as we can see, there is no empirical basis for assuming that individual quantum systems are necessarily in pure states. Nor is there any evidence showing that density matrices necessarily have to be interpreted as representing ensembles.  In fact, it seems to be accepted in the quantum information community that the state of an individual system should be represented, in certain circumstances, by a higher-rank density matrix. This can happen, for example, if the system is entangled with another system and the state of the composite system is pure, in which case the state of the first system is obtained by taking the reduced density matrix of the system as a whole, where we trace out the degrees of freedom associated with the second system. It thus seems reasonable  to take matters a step further and drop altogether the assumption that individual systems are necessarily in pure states. It also seems reasonable to drop the assumption that statistical ensembles play a fundamental role in the theory. In our approach, therefore, we make no use of frequentist thinking, and we avoid reference to observers, measurements, and ensembles. We regard state reduction as an entirely {\sl objective} phenomenon, and even in the case of an individual system we model the state  as a randomly evolving density matrix. 
We denote the density matrix process by $\{ \hat \rho_t \}_{t \geq 0} $, and we require  that  $\hat \rho_t$ should be nonnegative definite for all  $t$ and such that 
$\rm {tr} \,  \hat \rho_t =1$. The dynamical equation generalizing (\ref{pure dynamics})  then takes the following form:
\begin{definition}
\label{definition stochastic Lindblad}
We say that the state $\{ \hat \rho_t \}_{t \geq 0} $ of an isolated quantum system with Hamiltonian $\hat H$ satisfies an energy-driven stochastic master equation with parameter $\sigma$  if
\begin{eqnarray}
\label{density matrix process}
{\rm d}\hat \rho_t &=& -{\rm i}  \hbar^{-1} [ {\hat H}, \hat \rho_t ] {\rm d}t
+ \octa \sigma^2 \left( 2 \, \hat H  \hat \rho_t  \hat H - \hat H^2  \hat \rho_t - \hat \rho_t \hat H^2  \right) {\rm d}t \nonumber \\& & 
\quad \quad + \half
\sigma \left(  ({\hat H}-H_t) \hat \rho_t +  \hat \rho_t   ({\hat H}-H_t) \right) {\rm d}W_t , 
\end{eqnarray}
where $H_t = \rm tr \,\hat  \rho_t \hat H $.
\end{definition}

We take a moment to spell out some of the mathematical ideas implicit in the dynamics.  In accordance with the well-established Kolmogorovian outlook, we introduce a probability space $(\Omega, \mathcal F, \mathbb P )$ as the basis of the theory. We do not necessarily say in detail what the structure of this space is, but we assume that it is endowed with sufficient richness to support the various structures that we wish to consider. Thus $\Omega$ is a set on which we introduce a 
$\sigma$-algebra  $\mathcal F$ (no relation to the $\sigma$ above) and a probability measure $\mathbb P$. By an {\sl algebra} we mean a collection of subsets of $\Omega$ such that $\Omega \in \mathcal F$, $A \in \mathcal F$ implies $\Omega \backslash A \in \mathcal F$, $A \in \mathcal F$ and $B \in \mathcal F$ implies $A \cup B \in \mathcal F$. If for any countable collection of elements $A_i \in \mathcal F$, $i \in \mathbb N$, it holds that $\cup_{i\in \mathbb N} \, A_i \in \mathcal F$, then we say that $\mathcal F$ is a $\sigma$-{\sl algebra}.  The pair $(\Omega, \mathcal F)$ is called a {\sl measurable space}.  By a {\sl probability measure} on $(\Omega, \mathcal F)$ we mean a function $\mathbb P : \mathcal F \to [0, 1]$ satisfying $\mathbb P[\Omega] = 1$, 
and 
$\mathbb P[\cup_{i\in \mathbb N} \, A_i] = \sum_{i\in \mathbb N} \mathbb P[A_i]$ for any countable collection of elements $ A_i \in \mathcal F$,   $i \in \mathbb N$, such that $A_i \cap  A_j = \emptyset$ if $i \neq j$. A measurable space endowed with a probability measure defines a {\sl probability space}. 
A function $X : \Omega \to \mathbb R$ is said to be ${\mathcal F}$-{\sl measurable}, or measurable on $(\Omega, \mathcal F)$, if for all $A \in \mathcal B_{\mathbb R}$, where $\mathcal B_{\mathbb R}$ is the Borel $\sigma$-algebra on $\mathbb R$, it holds that
$\{ \omega \, : \, X(\omega) \in A \} \in \mathcal F$. Thus for each $A \in \mathcal B_{\mathbb R}$ we require
$X^{-1}(A) \in \mathcal F$. 
If $X$ is a measurable function on a probability space $(\Omega, \mathcal F, \mathbb P )$, we say that $X$ is a {\sl random variable}, and the associated distribution function is defined for $x \in \mathbb R$ by $F_X(x) = \mathbb P[X<x]$, where $ \mathbb P[X<x]$ denotes the measure of the subset 
$\{\omega \in \Omega : X(\omega) < x\}$. 

By a {\sl random process} on $(\Omega, \mathcal F, \mathbb P )$ we mean a family of random variables 
$\{X_t\}_{t\geq 0}$ parametrized by time. To formulate a theory of random processes some additional structure is required.  First we need the idea of a complete probability space. A $\sigma$-algebra $\mathcal F^P$ is said to be an {\sl augmentation} of the $\sigma$-algebra $\mathcal F$ with respect to $\mathbb P$ if  $\mathcal F^{\mathbb P}$ contains all subsets $B \subset \Omega$ for which there exist elements $A, C \in \mathcal F$ satisfying 
$A\subseteq B \subseteq C$ and $\mathbb P[C \backslash A] = 0$. If $\mathcal F^{\mathbb P} = \mathcal F$, we say that $(\Omega, \mathcal F, \mathbb P )$ is {\sl complete}. 
Next we need the idea of a  {\sl filtration} on $(\Omega, \mathcal F, \mathbb P )$, by which we mean a nondecreasing family $\mathbb F = \{\mathcal F_t\}_{t \geq 0}$ of sub-$\sigma$-algebras of $\mathcal F$. We say that a filtration $\mathbb F$ is {\sl right continuous} if for all $t\geq 0$ it holds that  
$\mathcal F_t = \mathcal F_{t^+} $ where
$\mathcal F_{t^+} = \cap_{u > t} \,  \mathcal F_u $.
If additionally we assume, as we do,  that for any $A \in \mathcal F$ such that 
$\mathbb P[A] = 0$ it holds that $A \in \mathcal F_0$, then we say that the filtered probability space
$(\Omega, \mathcal F, \mathbb P, \mathbb F)$ satisfies {\sl the usual conditions}.   
A random process $\{X_t\}$ is said to be {\sl adapted} to $\mathbb F$ if the random variable $X_t$ is $\mathcal F_t$-measureable for all $t \geq 0$.  We say that $\{X_t\}$ is right continuous if the sample paths $\{X_t(\omega)\}_{t\geq0}$ are right continuous for almost all $\omega \in \Omega$. 
By a {\sl standard Brownian motion} or {\sl Wiener process} on a filtered probability space  $(\Omega, \mathcal F, \mathbb P, \mathbb F )$  we mean a continuous, adapted process $\{W_t\}_{t\geq0}$ such that (a) $W_0 = 0$ almost surely, (b) $W_t - W_s$ is normally distributed with mean $0$ and variance $t - s$ for $t>s\geq0$, and (c) $W_t - W_s$ is independent of $\mathcal F_s$ for $t > s$. The filtration $\mathbb F$ may be strictly larger than that generated by the Brownian motion itself. The existence of processes satisfying these conditions is guaranteed by the following (Hida 1980, Karatzas \& Shreve 1986). Let $\Omega = C[0, \infty)$ be the space of continuous functions from $\mathbb R^+$ to $\mathbb R$. Each point $\omega \in \Omega$ corresponds to a continuous function $\{W_t(\omega)\}_{t\geq 0}$, and we write 
$\mathcal F = \sigma [ \{W_t \}_{t\geq 0} ]$ for the $\sigma$-algebra generated by $\{W_t \}_{t\geq 0}$. The $\sigma$-algebra generated by a collection $\mathcal C$ of functions $X : \Omega \to \mathbb R$ is defined to be the smallest $\sigma$-algebra $\Xi$ on $\Omega$ such that each function $X \in \mathcal C$ is $\Xi$-measurable. Then there exists a unique measure $\mathbb P$ on the $(\Omega, \mathcal F)$, called {\sl Wiener measure}, such that properties (a), (b) and (c) hold, and we take $\mathbb F$ to be the filtration $\{\mathcal F_t\}_{t\geq 0}$ generated by $\{W_t \}_{t\geq 0}$, defined by  $\mathcal F_t = \sigma [ \{W_s \}_{0\leq s \leq t} ]$ for each $t \geq 0$.  

In what follows we assume that $(\Omega, \mathcal F, \mathbb P, \mathbb F )$ satisfies the usual conditions. Equalities and inequalities for random variables are understood to hold $\mathbb P$-almost-surely. 
One checks by the use of Ito calculus that if $\hat \rho_t$ takes the form of a pure projection operator
\begin{eqnarray}
\label{pure state}
\hat \rho_t = \frac{  |\psi_t \rangle \langle{\psi}_t| } 
{\langle{\psi}_t|\psi_t\rangle} , 
\end{eqnarray}
then the stochastic Schr\"odinger equation (\ref{pure dynamics}) for the state-vector implies that the pure density matrix (\ref{pure state}) satisfies the stochastic master equation (\ref{density matrix process}). The relevant calculation is shown, for example, in sections 6.1-6.2 of Adler (2004). 
Since (\ref{density matrix process}) is a nonlinear stochastic differential equation, it does not immediately follow that (\ref{density matrix process}) should be applicable to general states rather than merely to pure states. Nevertheless, this is what we propose, and, as we shall see, the theory that follows from Definition  \ref{definition stochastic Lindblad} has many desirable properties, both physical and mathematical. For some purposes it is useful if we write equation (\ref{density matrix process}) in integral form, incorporating the initial condition explicitly. In that case we have
\begin{eqnarray}
\label{density matrix integral form}
\hat \rho_t &=& \hat \rho_0  -{\rm i}\, \hbar^{-1}\int_0^t [ {\hat H}, \hat \rho_s ] {\rm d}s
+ \octa \sigma^2 \int_0^t   \left( 2 \, \hat H  \hat \rho_s  \hat H - \hat H^2  \hat \rho_s - \hat \rho_s \hat H^2  \right) {\rm d}s \nonumber \\& & 
\quad \quad \quad + \half
\sigma \int_0^t  \left(  ({\hat H}-H_s) \hat \rho_s +  \hat \rho_s   ({\hat H}-H_s) \right) {\rm d}W_s \, .
\end{eqnarray} 
Then it follows, by taking the expectation of each side, which eliminates the term involving the stochastic integral, that the mean state of the system satisfies
\begin{eqnarray}
\label{expected density matrix integral form}
\langle  \hat \rho_t  \rangle = \hat \rho_0  -{\rm i}\, \hbar^{-1} \int_0^t [ {\hat H},  \langle  \hat \rho_s \rangle ] {\rm d}s 
+ \octa \sigma^2 \int_0^t   \left( 2 \, \hat H \langle   \hat \rho_s  \rangle  \hat H - \hat H^2   \langle \hat   \rho_s  \rangle - \langle  \hat \rho_s \rangle \hat H^2  \right) {\rm d}s.
\end{eqnarray} 
Here $\langle  \hat \rho_t  \rangle  = \mathbb E [\hat \rho_t]$, where ${\mathbb E} [\,\cdot \,]$ denotes expectation under $\mathbb P$.  One recognizes 
(\ref{expected density matrix integral form})
as the integral form of a master equation of the type derived by Lindblad (1976), Gorini et al (1976), and, in a different context, Banks et al (1984), and we have the following:
\begin{prop}
\label{Lindblad equation}
If the state of a quantum system satisfies the energy-driven stochastic master equation, then the mean state of the system satisfies a linear master equation of the form
\begin{eqnarray}
\label{expected density matrix differential form}
\frac { {\rm d}\langle  \hat \rho_t  \rangle } { {\rm d}t  } = -{\rm i} \, \hbar^{-1}[ {\hat H}, \langle  \hat \rho_t  \rangle ] 
+ \octa \sigma^2 \left( 2 \, \hat H  \langle  \hat \rho_t  \rangle \hat H - \hat H^2  \langle  \hat \rho_t  \rangle - \langle  \hat \rho_t  \rangle \hat H^2  \right) . 
\end{eqnarray}
\end{prop}
In the pure case, it is well known (see, for example, Gisin 1989) that if $|\psi_t \rangle$ satisfies (\ref{pure dynamics}) then the expectation of the corresponding pure density matrix, given by  
\begin{eqnarray}
\label{expected pure state}
\langle \hat \rho_t \rangle = {\mathbb E} \left[\frac{  |\psi_t \rangle \langle{\psi}_t| } 
{\langle{\psi}_t|\psi_t\rangle} \right],  
\end{eqnarray}
satisfies the autonomous stochastic differential equation (\ref{expected density matrix differential form}). This is not so obvious if one works directly with the dynamics of a state vector, but if one takes the stochastic master equation as the starting point then the linearity of the dynamics of $\langle \hat \rho_t \rangle$ is immediate.  
Proposition  \ref{Lindblad equation}
shows that in the generic situation where the density matrix is of rank greater than unity and follows the general nonlinear stochastic dynamics given by (\ref{density matrix process}), the associated mean density matrix  
$\langle  \hat \rho_t  \rangle$ still satisfies (\ref{expected density matrix differential form}).

We are thus led to postulate that the energy-driven stochastic master equation presented in Definition  \ref{definition stochastic Lindblad}, with a prescribed  initial state $\hat \rho_0$, 
characterizes the stochastic evolution of the state of a quantum system as reduction proceeds. In saying that we take the initial state as prescribed, we avoid for the moment entering into a discussion about how that can be achieved. Likewise, we avoid asking how one can determine what the initial state of the system is. It is meaningful to ask such questions, but we separate the problem of working out the consequences of the evolution of the state from the problem of working out what the state of the system is in the first place, or how to create a system in a given state. 
In Sections {\ref{Dynamic properties of the energy variance}, 
 \ref{Asymptotic properties of the variance} and \ref{Terminal value of the energy} below, we work out properties of  the energy-driven stochastic master equation. A number of the results obtained are generalizations of corresponding results known to hold in the case when the state is pure.  In Proposition  \ref{EV PROP} we show that the expectation of the variance of the energy goes to zero in the limit as $t$ grows large.  In Proposition \ref{existence of E_infinity} we show that there exists a random variable $H_{\infty} = \lim_{t \to \infty} H_t$ taking values in the spectrum of the Hamiltonian such that we have
$
\mathbb E\, [H_{\infty}] = {\rm tr} \, {\hat\rho}_0 \, {\hat H}$ and
$ {\rm Var}\, [H_{\infty}] = {\rm tr} \, {\hat\rho}_0 \, {\hat H^2} - ( {\rm tr} \, {\hat\rho}_0 \, {\hat H})^2.
$
The proofs of Propositions \ref{EV PROP} and  \ref{existence of E_infinity} generalize arguments appearing in Hughston (1996). 
In Section V we present a derivation of the Born rule for general states, summarized In Proposition  \ref{Born rule proposition}, extending arguments of Ghiradi et al (1990), Adler \& Horwitz (2000), and Adler et al (2001). 
In the case of a degenerate Hamiltonian, the reduction leads for a given outcome to the associated L\"uders state. 
Then in Sections VI and VII we proceed to construct a general solution of the energy-driven stochastic master equation using techniques of nonlinear filtering theory. Here we extend results known for the dynamics of pure states (Brody \& Hughston 2002). The solution, which takes the form of a completely positive stochastic map, is obtained by the introduction of a so-called {\sl information process} $\{\xi_t\}_{t\geq0}$ defined by $\xi_t = \sigma t H + B_t$ where the random variable $H$ takes values in the spectrum of the Hamiltonian operator, and $\{B_t\}_{t\geq0}$ is an independent Brownian motion. We show that it is possible to construct the processes $\{\hat \rho_t\}_{t\geq0}$ and $\{W_t\}_{t\geq0}$ in terms of $\{\xi_t\}_{t\geq0}$ in such a way that $\{\hat \rho_t\}_{t\geq0}$ satisfies the energy-driven stochastic master equation and $\{W_t\}$ is a standard Brownian motion on $(\Omega, \mathcal F, \mathbb P, \mathbb F )$, where $\mathbb F$ is the filtration generated by $\{\xi_t\}$. The results are summarized in Propositions  \ref{solution1} and \ref{solution2}. Then we introduce the notion of a {\sl potential} and in Propositions \ref{solution3} and \ref{state process form} we show that the decoherence of the density matrix  can characterized in a rather natural way by the fact that its off-diagonal terms are potentials. Section VIII concludes. 

\section{Dynamic properties of the energy variance}
\label{Dynamic properties of the energy variance}

\noindent We proceed to show that many of the important properties of the pure state dynamics (\ref{pure dynamics}) carry forward to the general state dynamics (\ref{density matrix process}). First, one can check that the trace of $\hat \rho_t$ is preserved under (\ref{density matrix process}). Thus, if $\rm tr \, \hat \rho_0 = 1$ then equation 
(\ref{density matrix integral form}) implies that $\rm tr \, \hat \rho_t = 1$ for all $t >0$. Next, one can check that the energy expectation process 
$\{H_t\}_{t \geq 0}$ defined by $H_t = {\rm tr} \, {\hat\rho}_t \, {\hat H} $
is a {\sl martingale}.
In fact, even in the pure case the result can be obtained rather more directly by use of  (\ref{density matrix process}) 
than (\ref{pure dynamics}), for if we transvect each side of (\ref{density matrix process})  with $\hat H$ and take the trace
we are immediately led to the following dynamical equation for the energy: 
\begin{eqnarray}
 \label{energy dynamics}
{\rm d}H_t = \sigma V_t \, {\rm d}W_t .
\end{eqnarray}
Here we have written $V_t = {\rm tr}\, \hat \rho_t \, ({\hat H}-H_t)^2$
for the variance of the energy.   
Thus we have
\begin{eqnarray}
\label{energy integrated}
H_t = H_0 + \sigma \int_0^t  V_s \,{\rm d}W_s.
\end{eqnarray}
Since the expectation and the variance of the energy are bounded random variables, it follows  from (\ref{energy integrated}) that $\{H_t\}_{t \geq 0}$ is a martingale. Letting 
${\mathbb E}_t[\, \cdot \,] = {\mathbb E} [\, \cdot \, | \, \mathcal F_t]$ denote conditional expectation with respect to 
$\mathcal F_t$, we have
$
{\mathbb E}_s [H_t] = H_s 
$
for $0 \leq s \leq t$. The martingale property represents conservation of energy in a conditional sense. This property is known to be satisfied by the energy expectation process in the case of a pure state, and we see that the martingale property holds more generally in the case of a mixed state governed by the energy-driven stochastic master equation. A further calculation shows that 
\begin{eqnarray}
 \label{variance dynamics}
{\rm d}V_t = -\sigma^2 V_t^2 \,{\rm d}t + \sigma \beta_t \,{\rm d}W_t,
\end{eqnarray}
where $\{\beta_t\}_{t \geq  0}$ denotes the so-called energy skewness process, defined by
\begin{eqnarray}
\beta_t = {\rm tr} \,\hat \rho_t \, ({\hat H}-H_t)^3.
\end{eqnarray}
The dynamical equation (\ref{variance dynamics}) can be obtained as follows. Write the variance in the form 
\begin{eqnarray}
V_t = {\rm tr}\, \hat \rho_t \, {\hat H}^2 - H_t^2.
\end{eqnarray}
The dynamics of the  term  
${\rm tr}\, \hat \rho_t \, {\hat H}^2$ can be worked out by transvecting each side of equation (\ref{density matrix process}) with ${\hat H}^2$. The dynamics of the second term can be deduced by applying Ito's lemma to $H_t^2$ and using (\ref{energy dynamics}). The two results combined give  
(\ref{variance dynamics}). 

The stochastic equation satisfied by the variance of the Hamiltonian in the case of a general state has the same form that it has in the pure case. 
In the pure case (\ref{variance dynamics}) implies that the variance tends to zero asymptotically, and thus that the state evolves to an energy eigenstate. We shall show that the argument carries through to the case of a general initial state. 
That is to say, for any initial state the result of the evolution given by (\ref{density matrix process}) is an energy eigenstate.  
By an energy eigenstate with energy $E$ we mean a state $\hat \rho$ such that 
$\hat H \hat  \rho = E \hat  \rho$.
If the Hamiltonian is nondegenerate, then the energy eigenstates are pure states.  In the case of a degenerate Hamiltonian,  the situation is more complicated. If the outcome of the collapse is an eigenstate with energy $E_r$, then it can be shown that the state that results is the so-called L\"uders state given by outcome of the L\"uders (1951) projection postulate associated with that energy and the given initial state (Adler et al 2001).  

\begin{definition} Let $\hat P_r$ denotes the projection operator onto the Hilbert subspace $\mathcal H_r$ consisting of state vectors that are eigenstates of $\hat H$ with eigenvalue $E_r$. Then for any initial state $\hat \rho_0$  the associated  L\"uders state  $\hat L_r$ is defined by
\begin{eqnarray}
\hat L_r =  \frac
{
\hat P_r  \,  \hat \rho_0 \,  \hat P_r
}
{
{\rm tr} \,    \hat \rho_0 \,  \hat P_r  
}.
 \end{eqnarray}
 \end{definition}
If the Hamiltonian is degenerate, and if the initial state is pure, then the final state will be pure. On the other hand, if the initial state is impure, then the final state need not be pure, and in general will be impure. 

To show that collapse to an energy eigenstate occurs as a consequence of (\ref{variance dynamics}) for a general initial state, we establish the following, which is known to hold for pure states: 
\begin{prop} Let $\{\rho_t\}$ satisfy the energy-driven stochastic master equation. Then 
the expectation of the variance of the Hamiltonian vanishes asymptotically:
\begin{eqnarray} 
\label{limit of EV}
\lim_{t \rightarrow \infty} {\mathbb E}[V_t] = 0.
\end{eqnarray}
\label{EV PROP}
\end{prop}
{\bf Proof} \,
We integrate  (\ref{variance dynamics}) to obtain
\begin{eqnarray}
\label{variance integrated}
V_t = V_0 -\sigma^2 \int_0^t V_s^2 \,{\rm d}s + \sigma \int_0^t  \beta_s \,{\rm d}W_s.
\end{eqnarray}
The integrals are defined since the variance and the skewness are bounded. Since the drift in 
(\ref{variance dynamics}) is negative, we see that ${\mathbb E}_s [V_t] \leq V_s$ for $0 \leq s \leq t$ and hence that $\{V_t\}_{t\geq 0}$ is a {\sl supermartingale}.  
Taking the unconditional expectation on each side of (\ref{variance integrated}), we have
\begin{eqnarray}
\label{expected variance}
{\mathbb E} [V_t] = V_0 -\sigma^2 {\mathbb E}\left[ \int_0^t V_s^2 \,{\rm d}s \right], 
\end{eqnarray}
which shows that ${\mathbb E}[V_t] $ decreases as $t$ increases, and hence that 
$\lim_{t \rightarrow \infty} {\mathbb E}[V_t]$ exists. 
 We say that an $\mathbb R$-valued random process $\{X_t\}_{t \geq 0}$ on a probability space 
$(\Omega, \mathcal F, \mathbb P)$ is $\sl measurable$ if for all $A \in \mathcal B_{\mathbb R}$, where $\mathcal B_{\mathbb R}$ is the Borel $\sigma$-algebra on $\mathbb R$, it holds that
\begin{eqnarray}
\{ (\omega, t) \, : \, X_t(\omega) \in A \} \in \mathcal F \times \mathcal B_{\mathbb R^+},
\end{eqnarray}
where $\mathcal B_{\mathbb R^+}$ denotes the Borel $\sigma$-algebra on the positive ``time axis" 
$\mathbb R^+ = [0, \infty)$. A sufficient condition for a process to be measurable is that it should be right continuous. 
Then one has the following (Liptser \& Shiryaev 1975): 

\vspace{3mm}
\noindent {Fubini's theorem}.\,
{\it If a process}  $\{X_t\}_{t\geq 0}$ {\it is measurable and} $\int_{S} \mathbb E [ \, | X_t | \, ] \, {\rm d}t < \infty$ {\it for some} 
$S \in \mathcal B_{\mathbb R^+}$,  {\it then} $\int_{S} \mathbb  | X_t | \,  {\rm d}t < \infty$ {\it almost surely and}
\begin{eqnarray}
\mathbb E \left [ \int_{S} X_t \, {\rm d}t \right ] = \int_{S} \mathbb E [ X_t ] \, {\rm d}t.
\end{eqnarray}
As a consequence of Fubini's theorem, we can interchange the order of the expectation and the integration on the right side of  (\ref{expected variance}) to obtain
\begin{eqnarray}
{\mathbb E} [V_t] = V_0 -\sigma^2  \int_0^t {\mathbb E}\left[V_s^2 \right] \,{\rm d}s , 
\end{eqnarray}
from which it follows that 
\begin{eqnarray}
\frac { {\rm d} {\mathbb E} [V_t] } { {\rm d} t}=  -\sigma^2 \, {\mathbb E}\left[V_t^2 \right] . 
\end{eqnarray}
Thus we can write
\begin{eqnarray}
\label{V bar dynamics}
\frac { {\rm d} {\mathbb E} [V_t] } { {\rm d} t}=  -\sigma^2 \, {\mathbb E} [V_t]^2 (1 + \alpha_t), 
\end{eqnarray}
where
\begin{eqnarray}
 \alpha_t = \frac{1}{{\mathbb E} [V_t]^2} {\mathbb E} [ (V_t - {\mathbb E} [V_t])^2 ], 
\end{eqnarray}
and we note that $\alpha_t$ is nonnegative. If we set 
$
\gamma_t = \int_0^t \alpha_s {\rm d}s, 
$
we can integrate (\ref{V bar dynamics}) to obtain
\begin{eqnarray}
{\mathbb E}[V_t]  = \frac {V_0} {1 + V_0\, \sigma^2 (t + \gamma_t)}.
\end{eqnarray}
Since $\gamma_t$ is nonnegative, we have 
\begin{eqnarray}
{\mathbb E}[V_t]  \leq \frac {V_0} {1 + V_0\, \sigma^2 t},
\end{eqnarray}
and this gives (\ref{limit of EV}).  
\hspace*{\fill} $\square$  

\vspace{3mm}

\section{Asymptotic properties of the variance}
\label{Asymptotic properties of the variance}
\noindent As a consequence of (\ref{limit of EV}) one deduces that the energy variance vanishes as $t$ goes to infinity.  More precisely, it holds that $V_{\infty} = 0$ almost surely. To see this, we need to show that the limit
$V_{\infty} = \lim_{t \rightarrow \infty} V_t$
exists, in an appropriate sense, and then we need to show that the order of the limit and the expectation in (\ref{limit of EV}) can be interchanged. 
If both of these conditions hold, then we conclude from (\ref{limit of EV}) that $V_{\infty} = 0$. Now, when we ask whether a limit exists, we are not asking whether the result is finite or not. Limits, if they exist, are allowed to be infinite. The question is one of convergence. Moreover, even if a random process converges,  that does not imply that the resulting function on $\Omega$ to which the process converges is a random variable (that is to say, a measurable function). So the question is whether there exists a random variable $V_{\infty}$ 
to which the variance process converges for large $t$ with probability one. If the answer is yes, then one can ask whether the interchange of limit and expectation is valid, and if so then we are able to conclude that the result of the collapse process is a state of zero energy variance and hence an energy eigenstate. 

To show that (\ref{limit of EV}) implies $V_{\infty} = 0$ almost surely, we use the {\sl martingale convergence theorem}. There are various versions of this theorem, and  it will be sufficient to have at hand the version that follows below (Protter 2003).  First we introduce some additional terminology. We fix a probability space and let $p \in \mathbb R$ satisfy $p \geq 1$. 
A random process  $\{X_t\}_{t\geq 0}$ is said to be bounded in $\mathcal L^p$  if 
\begin{eqnarray}
\sup_{0 \leq t < \infty} {\mathbb E}[\, | X_t | \, ^p\,] < \infty.
\end{eqnarray}
As usual, by the supremum we mean the least upper bound. A random process  $\{X_t\}_{t\geq 0}$ is said to be right-continuous if it holds almost surely that
$\lim_{\epsilon \to 0} X_{t + \epsilon} = X_t$ for all $t\geq 0$. Then we have: 

\vspace{3mm}
\noindent { Martingale convergence theorem}.\,
{\it If a right-continuous supermartingale 
$\{X_t\}_{t\geq 0}$ is bounded in $\mathcal L^1$ then $\lim_{t \rightarrow \infty} X_t$ exists almost surely and defines a random variable $X_{\infty}$ satisfying ${\mathbb E} [|X_{\infty}| ]< \infty$. }
\vspace{-1mm}

\noindent Note that in asserting that $\lim_{t \rightarrow \infty} X_t$ exists almost surely we mean that $ \limsup_{t \rightarrow \infty} X_t (\omega) = \liminf_{t \rightarrow \infty} X_t (\omega)$ for all $\omega \in \Omega'$ for some set $\Omega' \in \mathcal F$ such that $\mathbb P [\Omega'] = 1$, and  that  there exists a random variable $X_{\infty} $ such that $X_{\infty}(\omega) = \lim_{t \rightarrow \infty} X_t(\omega)$  for all $\omega \in \Omega$ apart from a set of measure zero.

As we shall see, the martingale convergence theorem is just the tool one needs in order to show that the energy variance process converges to zero. In particular, since the energy variance is bounded for all $t \geq 0$, we have 
$\sup_{0 \leq t < \infty} {\mathbb E}[\, | V_t |^{\,p} \,] < \infty$ for all $p\geq 1$.
It follows by the martingale convergence theorem   that 
$V_{\infty} = \lim_{t \rightarrow \infty} V_t$
exists almost surely and that ${\mathbb E} [V_{\infty} ]< \infty$. 
To proceed further we make use of the following (see, e.g., Williams 1991):  

\vspace{3mm}
\noindent  {Fatou's lemma.}\,
{\it Let $\{Y_k\}_{k \in \mathbb N}$ be a countable sequence of nonnegative integrable random variables. Then
$\mathbb E\, [  \liminf_{k \rightarrow \infty} Y_{k}]  \leq  
   \liminf_{k \rightarrow \infty}  \mathbb E\, [ Y_{k}] $. }
\vspace{3mm}

\noindent If $\{t_k\}_{k \in \mathbb N} $ is a countable sequence of times such that $\lim_{k \rightarrow \infty} t_k = \infty$, then for any process  
$\{X_t\}_{t\geq 0}$ such that  $X_{\infty} = \lim_{t \rightarrow \infty} X_t$ exists it holds that
$ \lim_{k \rightarrow \infty} X_{t_k} = X_{\infty}$. Thus, in our case we have
$\lim_{k \rightarrow \infty} {\mathbb E}[V_{t_k}] = 0$ 
and $ \lim_{k \rightarrow \infty} V_{t_k} = V_{\infty} $. 
 We know that 
 if $ \lim_{k \rightarrow \infty} Y_{k}$ exists then it is equal to 
 $ \liminf_{k \rightarrow \infty} Y_{k}$. Then by Fatou's lemma we have $\mathbb E\, [  \lim_{k \rightarrow \infty} V_{t_k}]  \leq  \lim_{k \rightarrow \infty}  \mathbb E\, [ V_{t_k}] $. It follows that  $\mathbb E\, [V_{\infty} ] =0$ and hence $V_{\infty} = 0$ almost surely, since the variance is nonnegative.


\section{Terminal value of the energy}
\label{Terminal value of the energy}

\noindent Let 
$\rm{Spec}[\hat H]$ denote the spectrum of the Hamiltonian. Then we have the following result, which shows that $H_t$ and $V_t$ are given at each time $t \geq 0 $ respectively by the conditional mean and the conditional variance of the terminal value of the energy:

\begin{prop}
\label{existence of E_infinity}
There exists a random variable $H_{\infty}$ on $(\Omega, \mathcal F, \mathbb P)$ taking values in $\rm{Spec}[ \hat H]$ such that $H_t = {\mathbb E}_t [H_{\infty}] $ and
$V_t = {\mathbb E}_t [(H_{\infty} - {\mathbb E}_t [H_{\infty}] )^2] $.
\end{prop}
{\bf Proof} \,
Since $\{H_t\}_{t\geq 0} $ is bounded by the highest and lowest eigenvalues of $\hat H$, we have 
$\sup_{0 \leq t < \infty} {\mathbb E}[\, | H_t | \,] < \infty$
and hence by the martingale convergence theorem the random variable
$H_{\infty} = \lim_{t \rightarrow \infty} H_t$ exists and ${\mathbb E}[ H_{\infty}] < \infty$. A process $\{X_t\}_{t\geq 0}$ on a probability space $(\Omega, \mathcal F, \mathbb P )$ is said to be uniformly integrable if, given any $\epsilon > 0$ there exists a $\delta$ such that 
\begin{eqnarray}
{\mathbb E} \left[ \, | X_t | \,  \mathds 1 (\, | X_t | > \delta \,) \right] < \epsilon
\end{eqnarray}
 for all $t\geq0$, where $\mathds 1 (\, \cdot \,) $ is the indicator function. Let $\{M_t\}_{t\geq 0}$ be a right-continuous martingale on a probability space $(\Omega, \mathcal F, \mathbb P )$ with filtration $\{\mathcal F_t\}_{t\geq 0}$. Then it is known that the following conditions are equivalent: 
(i) there exists a random variable $M_{\infty}$ such that 
$\lim_{t \rightarrow \infty} {\mathbb E} [ | M_t - M_{\infty} |] = 0$; 
(ii)  there exists a random variable $M_{\infty}$ satisfying 
${\mathbb E}\,[ M_{\infty}] < \infty$ such that 
$M_t = {\mathbb E}_t \,[ M_{\infty}] $ for all $t\geq0$;
(iii)  $\{M_t\}_{t\geq 0}$ is uniformly integrable. 
Clearly, any bounded martingale is uniformly integrable.  Since $\{H_t\}_{t\geq 0} $ is bounded, we have 
\begin{eqnarray}
H_t = {\mathbb E}_t [H_{\infty}] , 
\end{eqnarray}
as claimed. 
We turn now to the variance, in connection with which we use the following.

\vspace{3mm}
\noindent {Monotone convergence theorem}.\,
{\it For any increasing sequence $\{Y_k\}_{k \in \mathbb N}$ of nonnegative integrable random variables such that  $\lim_{k \rightarrow \infty} Y_{k} = Y_{\infty}$, where $Y_{\infty}$ is not necessarily integrable,  it holds that
$ \lim_{k \rightarrow \infty}  \mathbb E\, [ Y_{k}]  = \mathbb E\, [ Y_{\infty}] $}.
\vspace{3mm}

\noindent By use of (\ref{limit of EV}) and (\ref{expected variance}), together with the monotone convergence theorem, we deduce that
\begin{eqnarray}
 {\mathbb E}\left[ \int_0^{\infty} V_s^2 \,{\rm d}s \right] < \infty.  
\end{eqnarray}
Hence, it follows from (\ref{variance integrated}) that
\begin{eqnarray}
V_0 + \sigma \int_0^{\infty}  \beta_s \,{\rm d}W_s = \sigma^2 \int_0^{\infty}V_s^2 \,{\rm d}s. 
\label{V_0}
\end{eqnarray}
If we take a conditional expectation, we obtain
\begin{eqnarray}
V_0 + \sigma \int_0^{t}  \beta_s \,{\rm d}W_s = \sigma^2 \,
{\mathbb E}_t  \int_0^{\infty}V_s^2 \,{\rm d}s. 
\end{eqnarray}
Combining this relation with (\ref{variance integrated}) we deduce that
\begin{eqnarray}
V_t = \sigma^2 \, {\mathbb E}_t  \int_t^{\infty} V_s^2 \,{\rm d}s. 
\end{eqnarray}
Next we observe that as a consequence of (\ref{energy integrated}) we have
\begin{eqnarray}
H_{\infty} - H_t  =  \sigma \int_t^{\infty}   V_s \,{\rm d}W_s.
\end{eqnarray}
Taking the square of each side of this equation, forming the conditional expectation, and using the Ito isometry, we obtain
\begin{eqnarray}
{\mathbb E}_t \, (H_{\infty} - H_t )^2 
=  \sigma^2 \, {\mathbb E}_t  \int_t^{\infty} V_s^2 \,{\rm d}s ,
\end{eqnarray}
and therefore
\begin{eqnarray}
V_t = {\mathbb E}_t \, (H_{\infty}- {\mathbb E}_t H_{\infty} )^2,
\end{eqnarray}
as claimed.
 \hspace*{\fill} $\square$  \vspace{3mm}

The significance of this result is that the conventional expectation value $H_0$ of the observable $\hat H$ with respect to the initial state 
$\hat \rho_0$ is equal to the expectation of the terminal value of the energy on the completion of the reduction process. This may seem like a tautology, but it is not, since the statistical interpretation of the expectation value of an observable in quantum mechanics is an assumption, not a conclusion, of the theory. 

Likewise, we see that the conventional squared uncertainty $V_0$ is the variance of the terminal value of the energy on the completion of the reduction process. Again, the statistical interpretation of the squared uncertainty is an assumption in quantum mechanics, not a conclusion of the theory. But under the dynamics of the stochastic master equation these properties are deduced rather than assumed. 

The methods used  in the proof of Proposition \ref{existence of E_infinity} can be used to give an alternative derivation of the fact that $\lim_{t \rightarrow \infty} {\mathbb E}[V_t] = 0$ implies that $V_{\infty} = 0$ almost surely. We have already seen that this follows as a consequence of Fatou's lemma, but the same result can be obtained by use of the martingale convergence theorem. The proof is as follows. We observe that by the definition of the variance process we have
\begin{eqnarray}
 V_t = {\rm tr} \, \hat \rho_t \, \hat H^2 - H_t^2,
\end{eqnarray}
where $H_t =  {\rm tr} \, \hat \rho_t \, \hat H $. Writing $U_t = {\rm tr} \, \hat \rho_t \, \hat H^2$, we see that $\{U_t\}_{t \geq 0}$ is a bounded martingale. It follows by the martingale convergence theorem that 
$\{U_t\} \to U_{\infty}$, and as a consequence we have $\{V_t\} \to U_{\infty} - H_{\infty}^2$, from which it follows that
\begin{eqnarray}
 V_{\infty}  = U_{\infty}  - \left( H_0 + \sigma \int_0^{\infty}  V_s \,{\rm d}W_s \right)^2. 
\end{eqnarray}
Since $\mathbb E \, [ U_{\infty} ]  = U_0$, it follows by use of the Ito isometry that
\begin{eqnarray}
\mathbb E \, [ V_{\infty} ] 
= U_0 - H_0^2 - \sigma^2 \, \mathbb E \int_0^{\infty}  V_s^2 \,{\rm d}s. 
\end{eqnarray}
On the other hand, on account of (\ref{V_0}) we have
\begin{eqnarray}
V_0  = \sigma^2 \, {\mathbb E}  \int_0^{\infty}V_s^2 \,{\rm d}s. 
\end{eqnarray}
Since $U_0 - H_0^2 = V_0$, it follows that $\mathbb E \, [ V_{\infty} ] = 0$, and therefore $ V_{\infty} = 0$
almost surely.

\section{Derivation of the Born rule}
\label{Derivation of the Born rule}

\noindent The foregoing arguments show that the dynamic approach to reduction extends to the situation where the initial state of the system need not be pure.  The Born rule is another example of an assumption of quantum mechanics that can be derived from the stochastic master equation.  As before, for the given Hamiltonian let
 $ \hat  P_r $ denote the projection operator on to the Hilbert subspace of energy $E_r$. 
 Let the number of distinct energy levels be $D$. 

\begin{prop}
 \label{Born rule proposition}
Under the dynamics of the energy-driven stochastic master equation, with initial state $\rho_0$, the probability that the outcome will be a state with energy $E_r$ is given by
\begin{eqnarray}
\label{Born rule}
 {\mathbb P} \, [ H_{\infty} = E_r] = {\rm tr} \,  \hat  \rho_0  \hat  P_r.
 \end{eqnarray}
\end{prop}
\vspace{2mm}
{\bf Proof} \,
It is straightforward to check that the process 
 $\{ \pi_{rt} \}_{t \geq \infty}$ defined for each value of $ r = 1, 2, \dots, D$ by $\pi_{rt} =  {\rm tr} \,  \hat  \rho_t  \hat  P_r $ is a martingale. Thus we have 
$\pi_{rt} = {\mathbb E}_t\, [\pi_{r \infty} ]$ and hence
\begin{eqnarray}
{\rm tr} \,  \hat  \rho_0  \hat  P_r = 
{\mathbb E}\, [{\rm tr} \,  \hat  \rho_{\infty}  \hat  P_r ].
\end{eqnarray}
On the other hand, because the state reduces asymptotically to a random energy eigenstate, we know that  
\begin{eqnarray}
{\rm tr} \,  \hat  \rho_{\infty}  \hat  P_r = \mathds 1 ( H_{\infty} = E_r),
\end{eqnarray}
and since  
\begin{eqnarray} 
{\mathbb E} [ \mathds 1 ( H_{\infty} = E_r)] =
 {\mathbb P} [ H_{\infty} = E_r], 
 \end{eqnarray}
we are led to the Born rule (\ref{Born rule}).
 \hspace*{\fill} $\square$  \vspace{3mm}

It may seem tautological to assert that the probability of the outcome $E_r$ is  given by the trace of the product of the initial density matrix and the projection operator  $\hat  P_r$, but it is not.  In quantum mechanics, the Born rule is an assumption, part of the statistical interpretation of the theory. Physicists are on the whole quite comfortable with this assumption, but that does not change the fact that there is no generally accepted  ``derivation" of the Born rule as a probability law arising from within quantum theory itself. Indeed, it is one of the features of the energy-driven stochastic reduction model that a mathematically satisfactory explanation for this otherwise baffling aspect of quantum theory emerges. 

\section{Solution to stochastic master equation}
\label{solution to the stochastic master equation}

\noindent A solution to the energy-driven stochastic master equation (\ref{density matrix process}) can be written down as follows. We start afresh, and consider a finite-dimensional quantum system for which the Hamiltonian (possibly degenerate) is 
$\hat H$ and the initial state (which we regard as prescribed) is  $\hat \rho_0$. Let a probability space $(\Omega, \mathcal F, \mathbb P )$ be given, upon which we introduce a standard Brownian motion $\{B_t\}_{t\geq 0}$
and an independent random variable $H$ taking values in $\rm{Spec}[ \hat H]$ with the 
distribution $\mathbb P \,[ H = E_r] = {\rm tr} \,\hat  \rho_0  \hat  P_r$, where 
$ \hat  P_r $ denotes the projection operator on to the Hilbert subspace of energy $E_r$. Then we introduce a so-called  {\sl information process} on $(\Omega, \mathcal F, \mathbb P )$ denoted $\{\xi_t\}_{t\geq 0} $, defined by 
\begin{eqnarray}
\label{information process}
\xi_t = \sigma t H + B_t.
\end{eqnarray}
Thus $\{\xi_t\}$ takes the form of a Brownian motion with a random drift, the rate of drift being determined by the random variable $H$ and the parameter $\sigma$. Processes of this type arise in the theory of stochastic filtering (Wonham 1965, Liptser \& Shiryaev 2000). In the language of filtering theory one refers to $H$ as the {\sl signal}, $B_t$ as the {\sl noise}, and $\xi_t$ as the {\sl observation}. Of course, the notion of observation as it is understood in the context of filtering theory has no immediate connection with the notion of observation as it is usually understood in quantum mechanics. Nevertheless, the ideas that have been developed in filtering theory are rather suggestive, so it is worth keeping the associated terminology in mind as we proceed. Loosely speaking, one can think of ``that which has been observed" in the context of filtering theory as equivalent to ``that which has irreversibly manifested itself in the world" in the context of a physical theory.  Now, let $\{\mathcal F_t\}_{t \geq 0}$ denote the filtration generated by $\{\xi_t\}_{t\geq 0}$. We have the following. 
\vspace{2mm}

\begin{prop}
\label{solution1}
Let  the operator-valued  process $\{\hat K_t\}_{t\geq 0} $ be defined by
\begin{eqnarray}
\hat K_t = \exp \left [- {\rm i \, \hbar^{-1}}\hat H t + \half \sigma \hat H \xi_t 
- \quarter \sigma^2 \hat H^2 t  \right ] .
\end{eqnarray}
Then the process $\{\hat\rho_t\}_{t\geq 0} $ defined by 
\begin{eqnarray}
\label{solution}
\hat\rho_t = \frac
{ \hat K_t \, \hat\rho_0 \, \hat K^*_t }  { {\rm tr} \,[ \hat K_t  \, \hat\rho_0 \,  \hat K^*_t ] }
\end{eqnarray}
has trace unity, is nonnegative definite, and satisfies a stochastic master equation of the form
\begin{eqnarray}
\label{SLE}
{\rm d}\hat \rho_t &=& -{\rm i} \, \hbar^{-1} [ {\hat H}, \hat \rho_t ] {\rm d}t
+ \octa \sigma^2 \left( 2 \, \hat H  \hat \rho_t  \hat H - \hat H^2  \hat \rho_t - \hat \rho_t \hat H^2  \right) {\rm d}t \nonumber \\& & 
\quad \quad + \half
\sigma \left(  ({\hat H}-H_t) \hat \rho_t +  \hat \rho_t   ({\hat H}-H_t) \right) {\rm d}W_t , 
\end{eqnarray}
where $ H_t = {\rm tr} \, \hat \rho_t  \hat H$ and the process 
$\{W_t\}_{t\geq 0} $ defined by
\begin{eqnarray}
\label{W}
W_t = \xi_t - \sigma \int_0^t H_s {\rm d} s 
\label{BM}
\end{eqnarray}
is an $\{\mathcal F_t\}$-Brownian motion.
\end{prop}

\vspace{2mm}
\noindent {\bf Remark} \, Here we look at the stochastic master equation from a new point of view. Instead of regarding $\{W_t\}_{t\geq 0}$ as an ``input" to the model, we regards $\{\xi_t\}_{t\geq 0}$ as the input. Then both
$\{\hat \rho_t\}_{t\geq 0}$ and $\{W_t\}_{t\geq 0}$ are defined in terms of $\{\xi_t\}_{t\geq 0}$, and together they satisfy equation
(\ref{SLE}).
\vspace{2mm}

\noindent {\bf Proof} \, Let us set
$
\Lambda_t =
{\rm tr} \,[ \hat K_t  \, \hat\rho_0 \,  \hat K^*_t ].
$
From the cyclic property of the trace we obtain
\begin{eqnarray}
\Lambda_t = {\rm tr} \,\hat\rho_0 \exp \left ( \sigma \hat H \xi_t 
- \half\sigma^2 \hat H^2 t  \right ).
\end{eqnarray} 
By Ito's lemma, along with 
$ ( {\rm d} \xi_t)^2 =  {\rm d}t$, which follows from (\ref{information process}),
we have
\begin{eqnarray}
 {\rm d} \Lambda_t
= \sigma \, {\rm tr} \, \hat\rho_0 \hat H  \exp \left ( \sigma \hat H \xi_t 
- \half\sigma^2 \hat H^2  t  \right )  {\rm d} \xi_t ,
\end{eqnarray} 
and therefore
$
\label{d lambda}
 {\rm d} \Lambda_t
=  \sigma \, H_t  \, \Lambda_t  \, {\rm d} \xi_t,
$
since
\begin{eqnarray}
H_t =  
\frac { {\rm tr} \, \hat\rho_0 \hat H  \exp \left ( \sigma \hat H \xi_t 
- \half\sigma^2 \hat H^2 t  \right ) } 
{ {\rm tr} \, \hat\rho_0  \exp \left ( \sigma \hat H \xi_t 
- \half\sigma^2 \hat H^2 t  \right ) }. 
\label{energy estimate}
\end{eqnarray} 
If we write (\ref{solution}) in the form 
\begin{eqnarray}
\hat\rho_t = \frac {1} {\Lambda_t} \,
 \hat K_t \, \hat\rho_0 \, \hat K^*_t , 
\end{eqnarray}
a straightforward calculation using the Ito quotient rule then gives (\ref{SLE}). To establish that the process 
$\{W_t\}_{t\geq 0} $ defined by (\ref{BM}) is an $\{\mathcal F_t\}$-Brownian motion under $\mathbb P$
we use the so-called  {\sl L\'evy criterion}. We need to show (i) that
$( {\rm d}W_t )^2 = {\rm d}t$, and (ii)  that 
$\{W_t\}_{t\geq 0} $ is an $\{\mathcal F_t\}$-martingale under $\mathbb P$.  

The first property follows immediately as a consequence of the Ito multiplication rules applied to (\ref{information process}) and (\ref{W}). To check that the second  property holds we need to verify for 
$ s \leq t$ that 
$
{\mathbb E} [W_t \, | \, \mathcal F_s ] = W_s .
$
Let
$\mathcal G_t$ denote the $\sigma$-algebra generated by $H$ and 
$\{\xi_u\}_{0\leq u \leq t} $. Then $\mathcal F_t$, which is generated by 
$\{\xi_u\}_{0\leq u \leq t} $ alone, is a sub-$\sigma$-algebra of $\mathcal G_t$, and  for all $t \geq 0$ we have the tower property of conditional expectation:
\begin{eqnarray}
\label{tower propoerty}
{\mathbb E} \,[ {\mathbb E} \,[  \, \cdot \,| \, \mathcal G_t ] \,| \, \mathcal F_t ]
 = {\mathbb E} \, [\, \cdot \,| \, \mathcal F_t ].
\end{eqnarray}
Now, by (\ref{W}) it holds that
\begin{eqnarray}
\label{expansion of W}
{\mathbb E} [W_t \, | \, \mathcal F_s ] = {\mathbb E} [\xi_t \, | \, \mathcal F_s ]  
- \sigma \int_0^{t} {\mathbb E} [H_u \, | \, \mathcal F_s ] \, {\rm d}u.
\end{eqnarray}
As for the first term on the right side of (\ref{expansion of W}), it follows from (\ref{information process}) that
\begin{eqnarray}
{\mathbb E} [\xi_t \, | \, \mathcal F_s ]  
&=& \nonumber {\mathbb E} [B_t \, | \, \mathcal F_s ]   
+ \sigma t \, {\mathbb E} [H \, | \, \mathcal F_s ]  
\\  \nonumber &=&
{\mathbb E} [B_t \, | \, \mathcal F_s ]   + \sigma t \, H_s 
\\  \nonumber &=&
{\mathbb E} \,[ {\mathbb E} \,[  \, B_t  \,| \, \mathcal G_s ] \,| \, \mathcal F_s ]
+ \sigma t \, H_s 
\\  \nonumber &=&
{\mathbb E} [B_s \, | \, \mathcal F_s ]  + \sigma t \, H_s 
\\  &=&
\xi_s + \sigma (t-s) \, H_s ,
\end{eqnarray}
where we use the tower property to go from the second to the third line. In the second term on the right side of (\ref{expansion of W}), we use the fact that $\{H_t\}$ is a martingale to deduce that
\begin{eqnarray}
\int_0^{t} {\mathbb E} [H_u \, | \, \mathcal F_s ] \, {\rm d}u
=
\int_0^{s} H_u \, {\rm d}u + \int_s^{t} H_s \, {\rm d}u
=
\int_0^{s} H_u \, {\rm d}u + (t - s) H_s.
\end{eqnarray}
Thus putting together the results for the two terms on the right side of (\ref{expansion of W}) we have
\begin{eqnarray}
{\mathbb E} [W_t \, | \, \mathcal F_s ]
=  \xi_s -  \sigma \int_0^{s} H_u \, {\rm d}u = W_s,
\end{eqnarray}
which is what we wished to show.
\hspace*{\fill} $\square$  


\section{Information filtration} 
\noindent The collapse property in the case of a general state admits a remarkable interpretation in the language of stochastic filtering. As before, let us write $\hat P_r$ ($r=1, \dots, D$) for the projection operator onto the Hilbert subspace 
$\mathcal H_r$ consisting of state vectors with eigenvalue $E_r$.  For any element $|a\rangle \in \mathcal H_r$ we have $\hat H \, |a\rangle = E_r  \,|a\rangle  $, and for the Hamiltonian we can write
\begin{eqnarray}
\hat H =  \sum_{r =1}^D E_r \, \hat P_r .
\end{eqnarray}
Therefore, if we set
$
\hat R_{nm\,t}=   \hat P_n  \,  \hat \rho_t \,  \hat P_m 
$
then for the diagonal terms we have
\begin{eqnarray}
\label{diagonal terms}
\hat R_{nn\,t}= \frac
{
\hat P_n  \,  \hat \rho_0 \,  \hat P_n
\exp \left [\sigma E_n \xi_t 
- \half \sigma^2 E_n^2 \,  t  \right ] 
}
{
\sum_{r=1}^D p_r \exp \left [\sigma E_r \xi_t 
- \half \sigma^2 E_r^2 \,  t  \right ] 
} ,
\end{eqnarray}
where
$p_r =  {\rm tr} \,    \hat \rho_0 \,  \hat P_r $.
\begin{prop}
\label{solution2}
For each n  the process $ \{ \hat R_{nn\,t} \}_{t\geq0}$
is a uniformly integrable martingale, given by
\begin{eqnarray}
\hat R_{nn\,t} = {\mathbb E} \,\left[ \mathds 1 (H = E_n) \, | \,  \mathcal F_t  \right] 
\frac
{
\hat P_n  \,  \hat \rho_0 \,  \hat P_n
}
{
{\rm tr} \,    \hat \rho_0 \,  \hat P_n  
},
 \end{eqnarray}
 where   
 \begin{eqnarray}
 \label{conditional expectation of indicator}
{\mathbb E} \,\left[ \mathds 1 (H = E_n) \, | \,  \mathcal F_t  \right]  = 
  \frac
{
p_n
\exp \left [\sigma E_n \xi_t 
- \half \sigma^2 E_n^2 \,  t  \right ] 
}
{
\sum_{r=1}^D p_r \exp \left [\sigma E_r \xi_t 
- \half \sigma^2 E_r^2 \, t  \right ] 
} .
 \end{eqnarray}
\end{prop}
\vspace{2mm}
Thus, $\{ \xi_t \}$ carries {\sl partial information} about the value of the random variable $H$, which is revealed as time progresses, leading asymptotically to the outcome 
\begin{eqnarray}
\hat R_{nn\,\infty} = \mathds 1 (H = E_n) \frac
{
\hat P_n  \,  \hat \rho_0 \,  \hat P_n
}
{
{\rm tr} \,    \hat \rho_0 \,  \hat P_n  
},
 \end{eqnarray}
which is the L\"uders state that results under the projection postulate in the standard theory as a consequence of an energy measurement, with the outcome $E_n$, given that the initial state is $ \hat \rho_0$. In the present context there is no measurement as such. Nevertheless, the final state of the reduction process is a L\"uders state.
For each value of $n$ the corresponding diagonal element of the density matrix at time $t$ is given by the conditional expectation of the indicator function $\mathds 1 (H = E_n)$ given the value of $\xi_t$.  

Next we present a probabilistic formulation of the fact that the state decoheres as reduction proceeds. For any operator $\hat O$ let us write $|\hat O| =( {\rm tr} \, \hat O \hat O^{\dagger})^{1/2}$. 

\begin{definition} 
By a potential on a filtered probability space $(\Omega, \mathcal F, \mathbb P, \mathbb F)$, we mean a strictly positive right-continuous supermartingale $\{ \pi_t \}_{t \geq 0}$ with the property that 
$\lim_{t \to \infty} \mathbb E [ \pi_t ] = 0$. 
\end{definition}
Then we have the following:

\begin{prop}
\label{solution3}
For each $n,m$ such that $n \neq m$  the process $ \{ \,  | \hat R_{nm\,t} | \,  \}_{t\geq0}$
is a potential.
\end{prop}

\noindent {\bf Proof} \,
Let $n,m$ be such that $n \neq m$. The off-diagonal matrix elements of the state then take the form 
\begin{eqnarray}
\hat R_{nm\,t} =  \hat P_n  \,  \hat \rho_0 \,  \hat P_m
  \frac
{
\exp \left [-{\rm i} \, \hbar^{-1} (E_n - E_m) \, t + \half\sigma (E_n + E_m) \xi_t 
- \quarter \sigma^2 (E_n^2 + E_m^2) \, t  \right ] 
}
{
\sum_{r=1}^D p_r \exp \left [\sigma E_r \xi_t 
- \half \sigma^2 E_r^2 \, t  \right ] 
}. 
 \end{eqnarray}
Thus we have
\begin{eqnarray}
\hat R_{nm\,t} =  \hat P_n  \,  \hat \rho_0 \,  \hat P_m \,  \exp \left [-{\rm i}\, \hbar^{-1} (E_n - E_m) \,t  \right] \,  \Phi_{nm\,t} ,
 \end{eqnarray}
where
\begin{eqnarray}
\Phi_{nm\,t}  =   
  \frac
{
\exp \left [ \half\sigma (E_n + E_m) \xi_t 
- \quarter \sigma^2 (E_n^2 + E_m^2) \, t  \right ] 
}
{
\sum_{r=1}^D p_r \exp \left [\sigma E_r \xi_t 
- \half \sigma^2 E_r^2 \, t  \right ] 
}, 
 \end{eqnarray}
and a calculation shows that
 
\begin{eqnarray}
\label{Phi}
\Phi_{nm\,t}  =   \Pi_{nm\,t}
\exp \left [
- \tfrac{1}{8} \sigma^2 (E_n - E_m)^2 \, t  \right ], 
 \end{eqnarray}
where 
\begin{eqnarray}
\Pi_{nm\,t} =  \frac
{
\exp \left [ \half\sigma (E_n + E_m) \xi_t 
- \tfrac{1}{8} \sigma^2 (E_n + E_m)^2 \, t  \right ] 
}
{
\sum_{r=1}^D p_r \exp \left [\sigma E_r \xi_t 
- \half \sigma^2 E_r^2 \, t  \right ] 
}. 
 \end{eqnarray}
We claim that for any $\lambda \in \mathbb R$ the process $\{ \mu_t \}_{t \geq 0}$ defined by 
\begin{eqnarray}
\mu_t = \frac
{
\exp \left [ \lambda \xi_t 
- \tfrac{1}{2} \lambda^2 t  \right ] 
}
{
\sum_{r=1}^D p_r \exp \left [\sigma E_r \xi_t 
- \half \sigma^2 E_r^2 \, t  \right ] 
}
 \end{eqnarray}
is a martingale. To see that this is so, note that by use of Ito's lemma, together with the relation 
${\rm d} \xi_t =  \sigma H_t {\rm d} t + {\rm d} W_t,$
we have
${\rm d} \mu_t = (\lambda - \sigma H_t) \mu_t  {\rm d} W_t,$
and thus
\begin{eqnarray}
\mu_t = \exp \left [ \int_0^t (\lambda - \sigma H_s) \, \rm d W_s - \tfrac{1}{2} \int_0^t (\lambda - \sigma H_s)^2 \, \rm d s  \right ].
 \end{eqnarray}
 Since 
 $\{ H_t \}_{t \geq 0}$
  is bounded, we deduce that 
  $\{ \mu_t \}_{t \geq 0}$
  is a martingale.  Therefore, 
  $ \{\Pi_{nm\,t} \}_{t \geq 0}$ is a martingale and  
 $ \{  \Phi_{nm\,t}    \}_{t\geq0}$ 
 is a supermartingale.  By (\ref{Phi}) one sees that
 \begin{eqnarray}
 \mathbb E \left[   \Phi_{nm\,t}  \right] = 
 \exp \left [
- \tfrac{1}{8} \sigma^2 (E_n - E_m)^2 \, t  \right ], 
 \end{eqnarray}
and hence 
   \begin{eqnarray}
\lim_{t \to \infty}  \mathbb E \left[ \,   \Phi_{nm\,t}  \, \right] = 0,
 \end{eqnarray}
so $ \{  \Phi_{nm\,t}    \}_{t\geq0}$ is a potential. Finally, we observe that
 \begin{eqnarray}
 \left | \hat R_{nm\,t} \right |  =  \left | \, \hat P_n  \,  \hat \rho_0 \,  \hat P_m \, \right |
\Phi_{nm\,t}, 
 \end{eqnarray}
from which we obtain
  \begin{eqnarray}
\lim_{t \to \infty}  \mathbb E \left[ \,  \left | \hat R_{nm\,t} \right | \, \right] = 0
 \end{eqnarray}
for $n \neq m$, which is what we wished to prove.  
\hspace*{\fill} $\square$  
\vspace{2mm}

Thus, the potential property of the off-diagonal terms of the density matrix in the energy representation captures the essence of what is meant by decoherence.  We see that the decay of the off-diagonal terms of the density matrix is exponential in time, and that the decay rate for any particular such term is proportional to the square of the difference of the associated energy levels.  
In fact, we can take the representation of  $\{  \Phi_{nm\,t}    \}_{t\geq0}$ as a potential a step further. A calculation making use of the Ito quotient rule shows that 
  \begin{eqnarray}
{\rm d} \Phi_{nm\,t} = - \tfrac{1}{8} \sigma^2 (E_n - E_m)^2 \, \Phi_{nm\,t} \, {\rm d} t +
\half  \sigma (E_n + E_m - 2H_t) \,  \Phi_{nm\,t} \, {\rm d} W_t.
 \end{eqnarray}
As a consequence, for each $n,m$ such that $n \neq m$ we have
  \begin{eqnarray}
  \label{Phi in integral form}
 \Phi_{nm\,t} = 1 - \tfrac{1}{8} \sigma^2 (E_n - E_m)^2 \int_0^t \Phi_{nm\,s} \, {\rm d} s +
\half  \sigma \int_0^t (E_n + E_m - 2H_s) \,  \Phi_{nm\,s} \, {\rm d} W_s.
 \end{eqnarray}
Taking the limit as $t$ goes to infinity and using the fact that  $\Phi_{nm\,\infty} = 0$ almost surely, we deduce that
  \begin{eqnarray}
1 +
\half  \sigma \int_0^{\infty} (E_n + E_m - 2H_s) \,  \Phi_{nm\,s} \, {\rm d} W_s 
= \tfrac{1}{8} \sigma^2 (E_n - E_m)^2 \int_0^{\infty} \Phi_{nm\,s} \, {\rm d} s.  
 \end{eqnarray}
Then by taking a conditional expectation we obtain
  \begin{eqnarray}
1 +
\half  \sigma \int_0^{t} (E_n + E_m - 2H_s) \,  \Phi_{nm\,s} \, {\rm d} W_s 
= \tfrac{1}{8} \sigma^2 (E_n - E_m)^2 \, \mathbb E_t \left [ \int_0^{\infty} \Phi_{nm\,s} \, {\rm d} s \right] ,  
 \end{eqnarray}
from which it follows by use of  (\ref{Phi in integral form}) that
  \begin{eqnarray}
 \Phi_{nm\,t} 
= \tfrac{1}{8} \sigma^2 (E_n - E_m)^2 \, \mathbb E_t \left [ \int_t^{\infty} \Phi_{nm\,s} \, {\rm d} s \right ].  
 \end{eqnarray}
This identity gives us a representation of $\{  \Phi_{nm\,t}    \}_{t\geq0}$ as a so-called type-$D$ potential. 
More explicitly, if we define an increasing process $\{A_{nm \, t} \}_{t\geq0}$ by setting
  \begin{eqnarray}
A_{nm \, t}
= \tfrac{1}{8} \sigma^2 (E_n - E_m)^2 \, \int_0^{t} \Phi_{nm\,s} \, {\rm d} s,  
 \end{eqnarray}
then we have
  \begin{eqnarray}
 \Phi_{nm\,t} 
=  \mathbb E_t \left [ A_{nm \, \infty} \right ] - A_{nm \, t},  
 \end{eqnarray}
which is the canonical form for a potential of type $D$ (Meyer 1966). Thus we arrive at the following: 
\begin{prop}
\label{state process form}
The state process under energy-driven stochastic reduction is of the form
\begin{eqnarray}
 \hat \rho_t = \sum_{n = 1}^D
 {\mathbb E}_t \left [\mathds 1 (H = E_n) \right ] \, 
\frac
{
\hat P_n  \,  \hat \rho_0 \,  \hat P_n
}
{
{\rm tr} \,    \hat \rho_0 \,  \hat P_n  
}
+\sum_{n,m = 1}^D \mathds 1_{n\neq m}
 \hat P_n  \,  \hat \rho_0 \,  \hat P_m \,  \exp \left [-{\rm i}\, \hbar^{-1} (E_n - E_m)\,t \right] \,  \Phi_{nm\,t}.
 \end{eqnarray}
 \end{prop}
The conditional expectation in the first term is given by (\ref{conditional expectation of indicator}) and the potential in the second term is given by (\ref{Phi}).  At time zero, the two terms combine to give the initial density matrix $\hat \rho_0$. As the collapse proceeds, the first term converges to the L\"uders state 
associated with the selected energy eigenvalue $E_n$, and the second term tails off to zero. 
It should be emphasized that if the initial state is impure, and if the Hamiltonian is degenerate, then the final state will in general also be impure.

\section{Conclusion}
\noindent  In our development of the dynamic reduction program we have taken the view that the state of a single system can be described by a density matrix that may or may not be pure. The initial state $\hat \rho_0$ is prescribed, and its value at time $t$ is given by the random density matrix $\hat \rho_t$. 
The model is understood as describing an ``objective" reduction process, so there are no observers in the theory in the usual sense.
All the same, one can ask what is known at time $t$, in the sense of what has manifested itself in the world (or, let's say, in the experimenter's laboratory) at that time. For this purpose it seems reasonable to adopt the view that the standard interpretation of the filtration $\{\mathcal F_t \}$ gives an adequate answer. This means that for any overall outcome of chance $\omega \in \Omega$, the value of any 
$\mathcal F_t$-measureable random variable $X_t$ will be ``known" or will have manifested itself at (or before) time $t$.  In particular, the value of $\hat \rho_t$ itself will be known at time $t$, as will the value of the information process $\xi_t$. Now, it is not quite meaningful to ask how $\xi_t$ can be measured in a theory in which there are no measurements. Nevertheless, we are forced to the conclusion that the theory only makes sense if $\xi_t$ is known (in the sense of having manifested itself) at time $t$. In fact, Diosi (2015) has arrived at what we believe to be in essence a similar conclusion, that stochastic reduction models only really make sense if the $\{\xi_t\}$ process can in some appropriate sense be monitored in real time. This is not the same thing as saying that the quantum system is being {\sl actively} monitored (in the sense of Diosi 1988, Barchielli \&  Belavkin  1991, Barchielli 1993,  Wiseman 1996, Wiseman \& Diosi 2001, Barchielli \& Gregoratti 2009), since the monitoring that takes place in such considerations is within a framework of standard quantum dynamics, and some form of {\it ad hoc} collapse is required as an additional assumption to make the infinitesimal collapses occur in response to the monitoring. But it may be that in a laboratory situation it is possible to monitor $\{\xi_t\}$, or equivalently $\{H_t\}$, in the passive sense implicit in the structure of the information filtration of the models that we have here described. The class of information-based models that can be developed by use of the filtering techniques discussed in Sections VI and VII  can be extended to a wider set of models, in which the underlying noise is not Brownian motion but rather a general L\'evy process. Such processes, like Brownian motion, have the property of being stationary with independent increments, but are not generally Gaussian and can be discontinuous. Providing that a condition is satisfied ensuring the existence of exponential moments, L\'evy trajectories are suitable for characterizing a wide and extraordinarily diverse family of noise processes (Brody, Hughston \& Yang 2013). The development of relativistic analogues of the models considered here remains an open problem, though it seems reasonable to conjecture that in the relativistic case the reduction process should lead to states for which the total mass and spin take definite values. 

\begin{acknowledgments}
\noindent This work was carried out in part at the Perimeter Institute for Theoretical Physics, which is supported by the Government of Canada through Innovation, Science and Economic Development Canada and by the Province of Ontario through the Ministry of Research, Innovation and Science, and at the Aspen Center for Physics, which is supported by National Science Foundation grant PHY-1066293. DCB acknowledges support from the Russian Science Foundation (project 16-11-10218). We  are grateful to S. L. Adler, T. Benoist, I. Egusquiza, S. Gao and  B. K. Meister for helpful comments and discussions. 
\end{acknowledgments}

\end{document}